\documentclass[]{aa}  
\usepackage{amsmath,bm}
\usepackage{graphicx}
\usepackage{txfonts}
\usepackage{cuted}
\usepackage{soul}
\newcommand{\unit}[1]{\ensuremath{\,\mathrm{#1}}}
\def\kms{$\rm{km~s}^{-1}$}
\newcommand{\bma}[1]{\mathbf{#1}}

\begin{document}

\title{Winking filaments due to cyclic evaporation-condensation}
   
\titlerunning{Winking filaments}

   \subtitle{}

   \author{Yuhao Zhou\inst{1}
          \and
          Xiaohong Li\inst{1}
          \and
          Jie Hong \inst{2}
          \and
          Rony Keppens \inst{1}
          }

   \institute{Centre for mathematical plasma-astrophysics (CmPA), KU Leuven,
              Celestijnenlaan 200B, 3001 Leuven\\
              \email{yuhao.zhou@kuleuven.be}
              \and
              School of Astronomy and Space Science, Nanjing University, Nanjing 210023, PR China\\
             }

   \date{}

 
  \abstract
   {
   Observations have {shown} that some filaments appear and disappear in the H$\alpha$ line wing images periodically.   {There have been no attempts to model these "winking filaments"   thus far.}
   }
   {
   The evaporation\discretionary{-}{-}{-}condensation mechanism is widely used to explain the formation of solar filaments. Here, we demonstrate, for the first time, how multi-dimensional evaporation\discretionary{-}{-}{-}condensation in an arcade setup invariably causes a stretching of the magnetic topology. We aim to check whether this magnetic stretching during cyclic evaporation\discretionary{-}{-}{-}condensation could reproduce a winking filament.
   }
   {
   We used our open-source code MPI-AMRVAC to carry out 2D magnetohydrodynamic simulations based on a quadrupolar configuration.
   A periodic localized heating{, which} modulates the evaporation\discretionary{-}{-}{-}condensation process, {was imposed} before{, during,} and after the formation of the filament.
   Synthetic H$\alpha$ and 304 \AA\, images were produced to compare the results with observations.
   }
   {
   For the first time, we {noticed} the winking filament phenomenon {in a simulation of the formation of on-disk solar filaments, which was in good agreement with observations.}
   Typically, the period of the {winking} is different from the period of the impulsive heating.
   A {forced} oscillator model explains this difference and fits the results well.
   A parameter survey is also done to look into details of the magnetic stretching phenomenon. 
   {We found that the stronger the heating or the higher the layer where the heating occurs, the more significant the winking effect appears.}
   }
   {}

   \keywords{Sun: Corona – Sun: magnetic fields – magnetohydrodynamics (MHD) – Sun: filaments, prominences – Sun: oscillations}

   \maketitle
%

\section{Introduction}
\label{sec1}
Solar prominences or filaments are cool and dense structures in the solar corona \citep{tand1995, vial2015}.
As the surrounding corona is tenuous, they can easily be observed in multiple wavelengths, such as H$\alpha$ and extreme ultraviolet (EUV). This is convenient for detecting waves or oscillations in the solar corona, for instance, as well as for examining how they are manifested in prominences \citep{arre2018, chen2020}.

Due to the magnetic {buoyancy}, heavy prominences can be suspended high up in the solar corona.
There have been several mechanisms proposed to explain how such dense and cool plasma can enter the solar corona \citep[see][for example]{mack2010}.
Meanwhile, many of these mechanisms have been demonstrated either by works of observation \citep[e.g.,][]{chae2003, berg2011, zou2016} or simulation 
\citep[e.g.,][]{an1988, kane2015, fan2018}.
The evaporation\discretionary{-}{-}{-}condensation mechanism, which has been studied {in 1D geometry in detail in a number of previous works} \citep[see e.g.,][]{anti1999, anti2000, xia2011},{ }is among these mechanisms.
{Later on, this approach was} extended to multi-dimensional magnetohydrodynamic (MHD) models \citep[see e.g.,][]{xia2012, xia2016, li2022}.

This evaporation\discretionary{-}{-}{-}condensation mechanism works in the following way.
Cool plasma in the chromosphere, mainly from the higher chromosphere to lower transition region (TR) heights, enters the solar corona through evaporation.
The heating that causes such evaporation is usually modeled as a parametrized localized {heating source. It is denoted as $H_{\mathrm{loc}}$ so that it may be} distinguished from the background heating, $H_{\mathrm{bgr}}$, which is used to balance the coronal radiative cooling and maintain a hot corona.
In the simulations, taking \cite{zhou2021} as an example,  the {following expression is used for} the background heating, $H_{\mathrm{bgr}}$:
\begin{equation}
H_{\mathrm{bgr}}=H_0\exp\left(-\left(y-y_0\right)/\lambda_{0}\right),
\label{eq11}
\end{equation}
where $y$ is the vertical coordinate perpendicular to the solar surface and $\lambda_0$ is the heating scale height.
Control parameters $y_0$ and $H_0$ serve to set {the magnitude of the background heating}.
{The expression in} Eq.~\ref{eq11} is typically used in simulations which lack detailed lower-lying convection aspects and/or neglect adequate treatments of the full radiative transfer equation.
Its justification is based on the observational fact that the total energy transported into the corona (per unit area) should be in the order of $\sim 2\times10^{5}\unit{erg}\unit{cm^{-2}}$ \citep{with1977, asch2001}.
Recently, \cite{brug2022} studied the effect of other observationally motivated forms of background heating \citep[see][for example]{mand2000}, especially on the in situ thermal-instability-driven formation process of prominence structures in flux ropes (while ignoring condensation-evaporation aspects).

The localized heating term $H_{\mathrm{loc}}$ usually takes the following form:
\begin{equation}
H_{\mathrm{loc}}=H_1\exp\left(-\left|y-y_1\right|^s/\lambda_1^s\right),
\label{eq12}
\end{equation}
where $s=1$ or 2.
Although {this expression is similar to that of $H_{\mathrm{bgr}}$}, in this case, $H_{\mathrm{loc}}$ behaves quite differently.
From previous parameter surveys \citep{xia2011, john2019, pelo2022}, successful {simulations of} evaporation-condensation have taken $H_{0}$ to be on the order of $\sim 10^{-5}$--$10^{-4}\unit{erg}\unit{cm^{-3}}$, while $H_1$ is typically larger by two to three orders {of magnitude ($H_{\mathrm{1}}\sim 10^{-2}$--$10^{-1}\unit{erg}\unit{cm^{-3}}$).}
{The} scale height, $\lambda_0$, has a typical value of $\sim 10^2$ Mm or even higher, making the heating in the corona nearly a constant.
While $\lambda_1$ is usually much smaller, making the heating decay very quickly when the height $y$ deviates from $y_1$.
That is why $H_{\mathrm{loc}}$ is called localized heating.
In { previous} 1D simulations, $\lambda_1$ could be taken from {a range} 1\%--10\% of the total length of the loop, {which is typically 2--20~Mm for a 200-Mm-long loop} \citep{mull2004, klim2010, xia2011, pelo2022}.
However, in multi-dimensional simulations, a {narrower} range of values $\sim 2$--3~Mm is usually adopted \citep{kepp2014, xia2016, zhou2020, li2022, jerv2022}.
It should be noted that although the magnitude parameter $H_1$ is larger than $H_0$ by orders of magnitude, the total energy of $H_1$ integrated throughout the chromosphere and corona could be even one order smaller than $H_0$. This could be the reason why we still have no strong observational evidence for this localized heating. 

The {observable manifestation of the $H_{\mathrm{loc}}$ occurrence} is, thus, still under debate.
For example, random footpoint motions can lead to small-scale reconnections \citep[nanoflares, as suggested by][]{park1988} and/or Ohmic dissipation \citep{pont2020}, more recently evidenced as ``campfires" \citep{berg2021}.
Alfvén waves generated at the photosphere can also cause wave energy leakage into the chromosphere or transition region, through a variety of mechanisms \citep[see, e.g.,][]{degr2002, vanb2014, hows2019}.
Torsional motions or tornado-like features may be responsible for energy transfer from the lower atmosphere \citep{su2012,wede2012}.
Small-scale energy events should in any case be ubiquitous, but are hard to detect directly in observations in short wavelength bands.
Recent observations {in} non-thermal meter-wave radio channels provide some supporting evidence \citep{shar2022}.

Apart from studying {formation of prominences, their oscillations} have also widely received attention in recent years \citep{ball2005, trip2009, arre2018, chen2020}.
Prominence oscillations are ubiquitous, and they have been observed since the 1930s \citep{dyso1930}.
At that time, filaments were found {appearing and disappearing} periodically in the H$\alpha$ line center and wings when their line-of-sight (LOS) velocity is high enough.
As a result, this phenomenon has also been  dubbed ``winking filaments'' and it has, in fact, been frequently observed during an earlier time when spectroscopic observations were more popular, due to the lack of spatial resolution for imaging observations \citep[e.g.,][]{hyde1966, rams1966}.
From these observations, it has been deduced that the period of prominence oscillations should be a sort of intrinsic property.

Starting from the beginning of this century, with the development of instruments, more and more winking filaments have been reported.
Some of the winking filaments were reported as by-products of Moreton waves or EUV waves.
For example, \citet{eto2002} observed a winking filament caused by a Moreton wave using H$\alpha$ line center and $\pm0.8$~\AA\, wings and, similarly, \citet{okam2004} found a winking filament triggered by an EUV wave.
Winking filaments could also result from lower atmosphere reconnections or coronal shocks \citep{isob2006, grec2014}.

\citet{gilb2008} proposed that the occurrence of the winking filament phenomenon requires {LOS velocities of at least $30$\,--\,$40\,\mathrm{km\,s}^{-1}$}, which is related to the intensity contrast between the filament and the background.
With only a few selected wavelengths available, instead of having high spectral resolution data, the LOS velocity could only be estimated using approximations.
For example, \citet{isob2006} observed a filament oscillation with a period of 2~hrs.
Although this was mainly a horizontal oscillation, they used the method mentioned in \citet{mori2003} {obtaining} LOS {velocities} of 20--30~\kms, which are much higher than the value of 4~\kms {for} the horizontal direction.
Similarly, \citet{gilb2008} observed a winking filament with a period of 29~min and a maximum LOS velocity of 41~\kms.
However, \cite{jack2013} revisited this event with another method and considered that the maximum LOS velocity is only few \kms. In Table~\ref{tb1}, we list {several parameters of winking filament obtained from} selected observations.


\begin{table*}[h]
\newcommand{\tablecell}[2]{\begin{tabular}{@{}#1@{}}#2\end{tabular}}
\centering
\caption{Some typical observed winking filaments in observations}
\label{tb1}
\begin{tabular}{c|c|c|c|c|c}
\hline
Literature & Wave band & Trigger & \tablecell{c}{Period \\ (min)} & position & \tablecell{c}{$v_{\mathrm{LOS}}$ \\ (\kms)} \\
\hline
\citet{eto2002} & H$\alpha$ & Moreton wave & 15 & near {limb} & - \\
\citet{okam2004} & H$\alpha$ & EIT wave & 20 & near center & - \\
\citet{isob2006} & H$\alpha$ & flux emergence & 120 & near {limb} & 20--30 \\
\citet{gilb2008} & H$\alpha$ \& 10830 \AA & Moreton wave & 29 & near {limb} & 41 \\
\citet{asai2012} & H$\alpha$ & Moreton wave & 15 & near {limb} & 50 \\
\citet{grec2014} & H$\alpha$ & coronal shock & 16 & near center & 15 \\
\citet{shen2014} & H$\alpha$ & EUV wave & 11--22 & near center & 6--14 \\
\hline
\end{tabular}
\end{table*}

We point out that according to the definition of a winking filament, only the LOS {component of velocity is responsible for the winking}.
For a filament near the center of the solar disk, this would be an oscillation basically perpendicular to the solar disk, which is in line with intuitive expectation.
However, for a filament near the {solar limb}, it should be predominantly an oscillation parallel to the solar disk.
As demonstrated by our previous 3D MHD simulations of oscillating prominences \citep{zhou2018}, the restoring force (and the associated period) for these two kinds of oscillations would be rather different.
Therefore, in the present work, the term "winking filament" only refers to an oscillation perpendicular to the solar disk. In reality, however, solar filaments are expected to be in a rather dynamic equilibrium state over most of their lifetimes \citep[see, e.g.,][]{berg2010, xia2016, berg2017, zhou2020, jenk2022a, jerv2022}.
Thus, all kinds of dynamics (perpendicular and {parallel}) are likely to be coupled with each other.

In {this work, we use the simulation to show} that the formation process of solar filaments by evaporation\discretionary{-}{-}{-}condensation mechanism can actually be responsible for the curious phenomenon of the winking filament. Our novel idea invokes an observationally justifiable temporal variability in the localized heating, and we show how the magnetic stretching in a 2D setup then leads to the winking in synthetic spectroscopic views.
The paper is organized as follows. In Sect.~\ref{sec2}, we discuss the setup of the simulation and the phenomenon of magnetic stretching; Sect.~\ref{sec3} presents a simulation of a winking filament. Section~\ref{sec4} gives our conclusion and a discussion.

\section{Evaporation\discretionary{-}{-}{-}condensation and magnetic stretching}
\label{sec2}
In this section, we use a 2D model to demonstrate that a magnetic stretching phenomenon is embedded in the traditional evaporation\discretionary{-}{-}{-}condensation mechanism.
This is a purely multi-dimensional effect that can never be studied with the restricted 1D fixed-field assumption that is often made in evaporation-condensation scenarios.

\subsection{Numerical setup}
The 2D magnetohydrodynamic (MHD) model used here is similar in setup with previous work \citep{kepp2014}.
The simulation box ranges from $-50$~Mm $<x<50$~Mm and $0<y<80$~Mm.
We used a 768$\times$768 uniform grid so that the resolution was 130~km $\times$ 104~km. 

The atmosphere used in our simulation is composed of pure {hydrogen}.
For the region below {a certain height} $y_c=2543$~km, we adopted the temperature profile $T(y)$ from the traditional VAL-C model \citep{vern1981}.
However, for the region above $y_c$, the following expression is used to extrapolate the semi-empirical VAL-C model so that the vertical heat flux is constant:
\begin{equation}
    T(y)={\left(7{F_\mathrm{c}}\left(y - {y_c}\right)/\left(2\kappa_\parallel\right) + {T_{\mathrm{tr}}}^{7/2}\right)^{2/7}}.
    \label{eq21}
\end{equation}
We set {the temperature at $y_c$, which is the typical temperature at the TR}, $T_{\mathrm{tr}}=4.47\times 10^5$~K so that the two profiles are continuously connected at $y_c$.
The constant vertical thermal conduction $F_\mathrm{c}$ is $2\times10^5\unit{erg}\unit{cm}^{-2}\unit{s}^{-1}$ and the Spitzer-type conductivity $\kappa_{\parallel}$ is $8\times10^{-7}{T^{5/2}}\unit{erg}\unit{cm}^{-1}\unit{s}^{-1}\unit{K}^{-1}$ in this work.
The atmosphere is partially ionized.
The ionization degree of {hydrogen} is approximated by a single function of temperature \citep{hein2015}.
The technical details will be given in the Appendix.
Then, together with {hydrogen number density at the bottom of the computational domain} $n_{\mathrm{Hb}}=9.45\times 10^{13}\unit{cm}^{-3}$, we can setup the hydrostatic equilibrium atmosphere. This specifies both the density and the pressure profile ({in 1D along the height}).

After setting up the atmosphere, we added a potential, quadrupolar field that has a magnetic dip located at the horizontal center of our 2D domain:
\begin{eqnarray}
    {B_x}  & =  & + {B_{\mathrm{p0}}}\cos \left( {\frac{{\pi x}}{{2{L_0}}}} \right){e^{ - \frac{{\pi y}}{{2{L_0}}}}} - {B_{\mathrm{p0}}}\cos \left( {\frac{{3\pi x}}{{2{L_0}}}} \right){e^{ - \frac{{3\pi y}}{{2{L_0}}}}},
    \label{eq22} \\
    {B_y} & = &  - {B_{\mathrm{p0}}}\sin \left( {\frac{{\pi x}}{{2{L_0}}}} \right){e^{ - \frac{{\pi y}}{{2{L_0}}}}} + {B_{\mathrm{p0}}}\sin \left( {\frac{{3\pi x}}{{2{L_0}}}} \right){e^{ - \frac{{3\pi y}}{{2{L_0}}}}}.
    \label{eq23}
\end{eqnarray}
We set $L_0=50$~Mm while $B_{\mathrm{p0}}$ is chosen to be 2~G in the demo case so that the plasma $\beta$ near the TR is approximately unity and the magnetic field strength near the magnetic dip is approximately 4--5~G.

The governing equations used in this purely 2D (all vectors have only $x,y$ components) simulation are as follows:
\begin{equation}
    \frac{{\partial \rho }}{{\partial t}} + \nabla  \cdot (\rho \bma{v}) = 0,
    \label{eq24}
\end{equation}
\begin{equation}
    \frac{{\partial (\rho \bma{v})}}{{\partial t}} + \nabla  \cdot \left(\rho \bma{vv} + {p_{\mathrm{tot}}}\bma{I} - \frac{{\bma{BB}}}{{{\mu _0}}}\right) = \rho \bma{g},
    \label{eq25}
\end{equation}
\begin{equation}
\begin{aligned}
    \frac{{\partial e}}{{\partial t}} + \nabla  \cdot &\left( {e\bma{v} + {p_{\mathrm{tot}}}\bma{v} - \bma{BB} \cdot \bma{v}} \right) \\&= \rho \bma{g} \cdot \bma{v} + \nabla  \cdot \left( {\bm{\kappa}  \cdot \nabla T} \right) - {n_\mathrm{e}}{n_\mathrm{H}}\Lambda \left( T \right) + H,
    \label{eq26}
\end{aligned}
\end{equation}
\begin{equation}
    \frac{{\partial {\bma{B}}}}{{\partial t}} + \nabla  \cdot (\bma{vB - Bv}) = \mathbf{0}.
    \label{eq27}
\end{equation}
Here, $\bma{g}=-{g_\odot}{r_\odot}^2/{({r_\odot}+y)^2}{\mathbf{\hat e}_y}$ is the gravitational acceleration and we have ${g_ \odot} = 274 \unit{m}\unit{s^{-2}}$, ${r_\odot}=691 \unit{Mm}$.
{The expression} $\nabla  \cdot \left( {\bm{\kappa}  \cdot \nabla T} \right)$ is the field-aligned Spitzer-type anisotropic thermal conduction where {$\kappa$ is tensor defined as} $\kappa=\kappa_{\parallel}\bm{\hat{b}}\bm{\hat{b}}$ and ${n_\mathrm{e}}{n_\mathrm{H}}\Lambda \left( T \right)$ is the optically thin radiative cooling term taken from the tables from \citet[][ for $T$<14000 K]{dalg1972} and \citet[][ for $T$>14000 K]{colg2008}.
For the region below $y=2.5$~Mm, this radiative cooling term is turned off because,  in the lower layers, we would need a more consistent radiative transfer treatment for the chromosphere.
The heating term, $H,$ is given later.
All the other symbols in the equations have their usual meanings. 

For boundary conditions, symmetric settings are adopted for $\rho$, $v_y$, $p,$ and $B_y$, while $v_x$ and $B_x$ are antisymmetric, on side boundaries.
All the variables are fixed on the bottom boundary except that $v_x$ and $v_y$ are asymmetric.
For the top boundary, $B_x$ and $B_y$ are extrapolated, $v_x$ and $v_y$ are asymmetric, while $\rho$ and $p$ are calculated according to a hydrostatic equilibrium assumption based on an extrapolated $T$.
Other numerical settings, including the scheme, adopted limiter and divergence cleaning method, are the same as in our previous work \citep{zhou2021}, except the transition region adaptive conduction (TRAC) method, which is not used in this work.
Simulations are done with our open source code MPI-AMRVAC\footnote{https://www.amrvac.org/}~\citep{xia2018}.

\subsection{Magnetic stretching}
We first relaxed the system {till} $t_{\mathrm{relax}}=214.7$~min to a thermodynamic equilibrium state.
During the relaxation, only a background heating term, $H=H_{\mathrm{bgr}}$, was imposed to balance the radiative cooling.
A time-dependent version of Eq.~\ref{eq11} is: 
\begin{equation}
H_{\mathrm{bgr}}\left(y, t\right)=R_{\mathrm{bgr}}\left(t\right)H_0\exp{\left(y/\lambda_0\right)},
\label{eq28}
\end{equation}
where $R_{\mathrm{bgr}}\left(t\right)=\max\left(5-12t/t_{\mathrm{relax}},1\right)$ is used to setup a steady background heating after $t=t_{\mathrm{relax}}/3$.
We chose $H_0$ to be $6\times10^{-5}\unit{erg}\unit{cm^{-3}}$ and $\lambda_0$ is 100~Mm.

\begin{figure*}
  \centering
  \includegraphics[width=16cm]{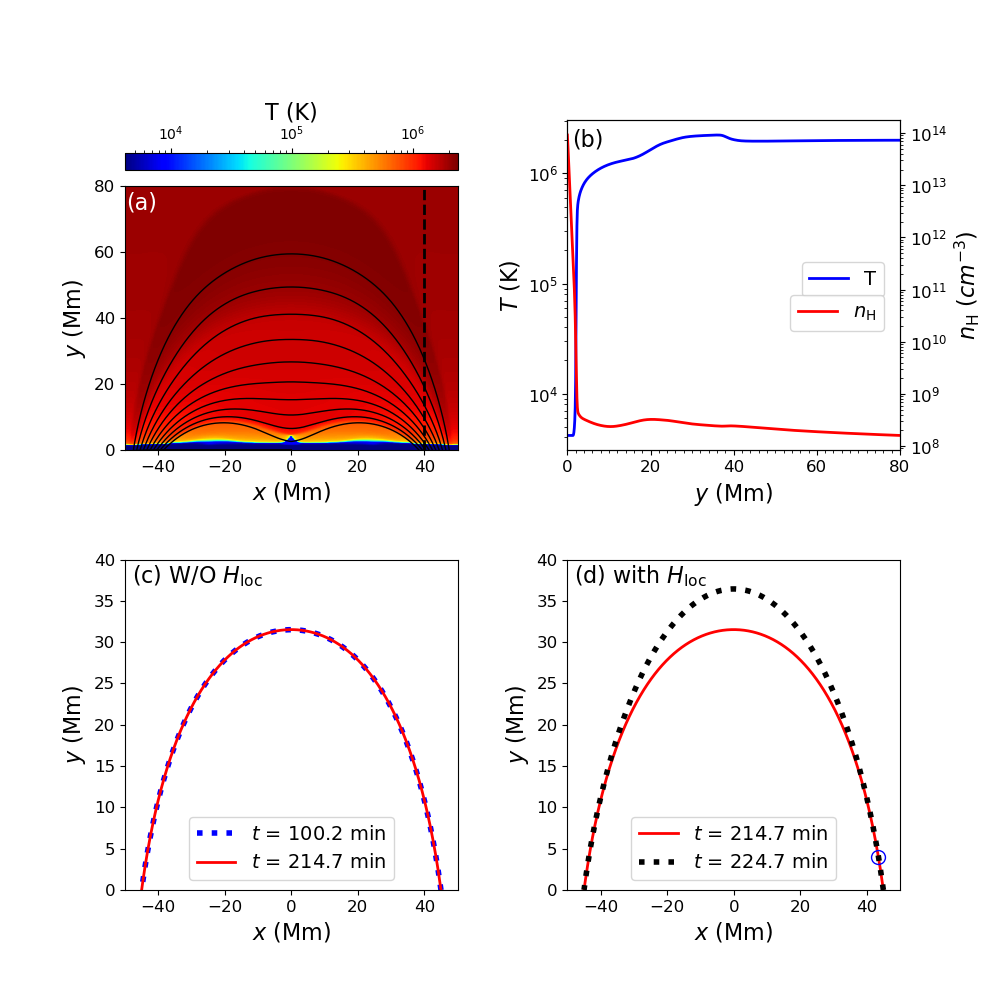}
  \caption{Initial condition and magnetic stretching. (a) Temperature distribution and magnetic configuration after relaxation. (b) Temperature and number density distributions along $x=40$~Mm, as indicated by the dashed line in panel a. (c) Selected magnetic field line at $t=100.2$ min and after relaxation. (d) Selected magnetic field line before and after we impose the localized heating.}
  \label{fig1}
\end{figure*}

The black solid lines in Fig.~\ref{fig1} {(a)} show the topology of the magnetic field with some selected field lines, at the end of the relaxation stage $t=214.7$~min.
The color scale shows the temperature distribution at this moment.
To show these distributions more clearly, a vertical slice is taken along the $y$-axis at $x=$40~Mm.
Temperature and number density (of {hydrogen}) distributions along this slice are shown in Fig.~\ref{fig1}{(b)}, with blue and red solid lines, respectively.
We can see that the TR is located a little bit higher than $y=2$~Mm.

Actually, the system is already in a relatively stable state after about 100~min, as shown in the later analysis.
However, we continue to relax the system for around 100 min to make a clear comparison between the relaxation stage and the further localized heating stage.
To show how a magnetic field line indeed no longer changes in the latter half of the relaxation phase, we select the field line which starts from $(x, y)=(45, 0)$~Mm (so that it ends at $(x, y)=(-45, 0)$~Mm due to its symmetry) and plot it in Fig.~\ref{fig1} {(c)}.
In that panel, we compare the trace of this field line at $t=100.2$~min (blue dashed line) and at $t=214.7$~min (red solid line).
Clearly, we can see that during the extended relaxation stage (without localized heating, $H_{\mathrm{loc}}$), this field line can stay at the same position for more than 100~min.

Then we start to impose the localized heating term $H_{\mathrm{loc}}$.
The localized heating term is a 2D version of Eq.~\ref{eq12} together with a modulating ramp function, $R_{\mathrm{ramp}}$, so that the localized heating can increase smoothly from zero after $t_{\mathrm{relax}}$:

\begin{eqnarray}
{H_{\mathrm{loc}}\left(x, y, t\right)} &=& {H_1}{R_{\mathrm{{ramp}}}}\left(t\right)H_x\left(x\right)H_y\left(y\right),
\label{eq29} \\
H_x\left(x\right) &=& \exp \left( { - \frac{{{{\left( {x - {x_r}} \right)}^2}}}{{{\sigma ^2}}}} \right) + \exp \left( { - \frac{{{{\left( {x - {x_l}} \right)}^2}}}{{{\sigma ^2}}}} \right),
\label{eq210} \\
H_y\left(y\right)  &=& \exp \left( { - \frac{{{{\left( {y - {y_1}} \right)}^2}}}{{\lambda _1^2}}} \right).
\label{eq211}
\end{eqnarray}

{Here,} $H_1=2 \times 10^{-2}~\unit{erg}\unit{cm^{-3}}\unit{s^{-1}}$, $y_1 = 4$ Mm and $\lambda _1 = 3.16$~Mm.
{In the $x-$direction, we use the parameters $x_l= -41.5$~Mm, $x_r= 41.5$~Mm, and $\sigma = 5.48$~Mm to ensure that the heating is concentrated at footpoints.}
We simply take a periodic sine function for the ramp function $R_{\mathrm{ramp}}$:
\begin{equation}
 R_{\mathrm{ramp}}=\max\left(\sin\left(2\pi\left(t-t_{\mathrm{relax}}\right)/P\right),0\right).
    \label{eq212}
\end{equation}
The period $P$ is taken as 20~min, a typical value listed in Table~\ref{tb1} {and also a typical value from the our previous simulation work on vertical oscillations \citep{zhou2018}.
Then the} first heating pulse will end in 10~min.

After the first 10~min, when $t=224.7$~min, as shown in Fig.~\ref{fig1} {(d)}, we can see that the field line we showed earlier gets stretched significantly (black dashed line), compared to its shape from 10 min before (red solid line).
Quantitatively, the apex of the loop rises from 3.15~Mm to 3.64~Mm, with an increment of 16\%.
The inferred velocity of such a rise during these 10 minutes is on average 8.2~\kms.

In previous works, such a vertical stretching of the magnetic field line is usually considered to be associated with eruptive events, typically erupting solar filaments and/or magnetic flux ropes \citep[see, for example,][]{chen2002, amar2014, jian2021}, in which plenty of energy is released.
However, in our numerical experiment shown above, even a small portion of gradual energy injection into the system can lead to a significant stretching.
We will look into the details of the physics here in Sect.~\ref{sec4}.

Our 2D MHD simulation shows that this kind of evaporation-induced field line stretching, where the increased thermal pressure gives rise to an overall inflated magnetic topology, should be ubiquitous on the Sun.
With such a natural multi-dimensional mechanism, we can now propose an alternative explanation for the winking filament phenomenon, that would be entirely driven by the popular evaporation\discretionary{-}{-}{-}condensation scenario for prominence formation.

\section{Winking filament}
\label{sec3}
In {Sect.~\ref{sec2}}, we have shown that the magnetic field lines in an arcade setup would be stretched during an evaporation\discretionary{-}{-}{-}condensation process.
However, we cannot actually see the configuration of magnetic field {lines} in observations.
Luckily{,} there are at least two ways {of how} to ``see'' this stretching indirectly by observing {the motion} of a filament in the arcade {that} is supported by the field line or by observing a possible wave generated by this stretching.
We go on to demonstrate that this can indeed induce the vertical oscillation of the filament, thereby leading to the winking filament.

As we did not yet have a filament in our setup at $t=224.7$~min, we continue our simulation from {the point where we stopped in Sect.~\ref{sec2}}.
We still adopt the periodic heating from Eq.~\ref{eq29} with the period of $P=20$~min.
As shown in \citet{john2019}, adopting different periods and durations of the added impulsive heating will lead to different scenarios involving thermal non-equilibrium evolutions and/or {condensations driven by thermal instabilities.}
For the typical period range listed in Table~\ref{tb1}, this influence should be minor and we expect to be able to form a large-scale prominence.

\begin{figure*}
  \centering
  \includegraphics[width=16cm]{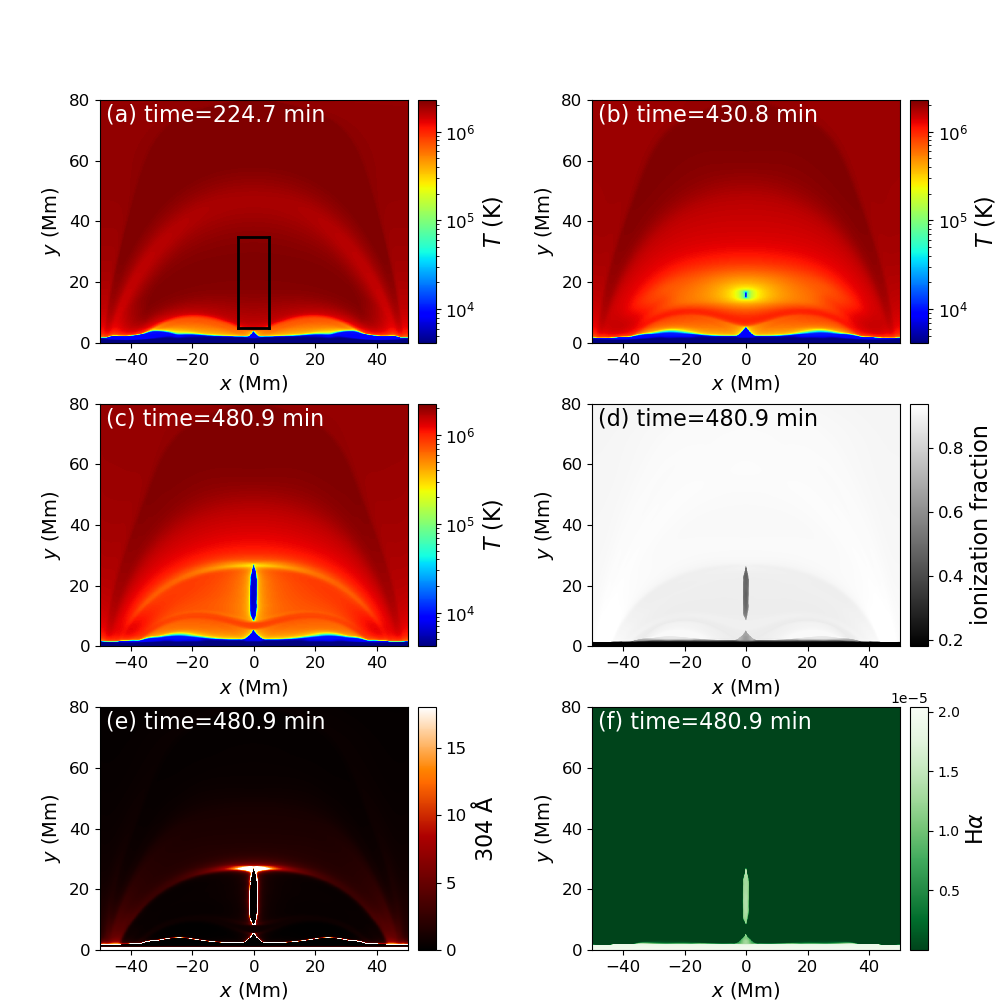}
  \caption{Formation of the filament. (a)-(c) Temperature distribution{s} during the formation of the filament at {the times} $t=224.7$~min, $t=430.8$~min {and} $t=480.9$~min. An animation showing {their} evolution is available {as online supplement of the paper.} (d) Ionization fraction distribution at $t=480.9$~min. (e) AIA 304~\AA\, synthetic image at $t=480.9$~min. (f) Approximate H$\alpha$ synthetic image at $t=480.9$~min. The units of the last two panels are arbitrary. See text for details.}
  \label{fig2}
\end{figure*}

Considering the condensation-evaporation mechanism, it is expected to cause a condensation, when localized heating is added for a limited period of time.
This first condensation grows larger and gradually forms a filament .
Figure~\ref{fig2} {(a-c)} shows the temperature evolution during the filament formation process in our simulation.
{In panel a, at} $t=224.7$~min, the average coronal temperature inside the black rectangle ($-5$~Mm $< x < 5$~Mm and 5~Mm $<y < 35$~Mm) reaches a maximum value of 2.23~MK, increased from the temperature of 1.55~MK before we imposed the localized heating.
After that, the temperature (inside the chosen rectangle) starts to drop gradually.
At around $t=430.8$~min, catastrophic cooling due to thermal instability happens so that a clear condensation starts to appear in the simulation domain ({panel} b).
Here, we define a condensation {as} cool material or filament as plasma below 14000~K, because {the values of} the ionization degree {listed in Table 1 in} \citet{hein2015} {were calculated for temperatures} from 6000~K to 14000~K.
Then, the area and mass of the cool filament increase gradually.
{Panel c} shows the temperature distribution at $t=480.9$~min, when we can clearly see a vertical sheet-like filament that formed.
From the attached animation, we can already clearly see the oscillation in the vertical direction.
At the same time, we can also see a periodic shrinking and expansion of the filament in the horizontal direction.

Fig.~\ref{fig2} (d) gives the distribution of the ionization fraction {for} the {time as is for} {panel} c.
{Because} the ionization fraction in \citet{hein2015} only ranges  from 0.17 to 0.94{,} we call {here} the plasma fully ionized when its ionization degree is near 0.94, instead of 1.
Most of the {hydrogen} within the condensation region is only partially ionized, as expected.
We can also see that most of the region within the outmost heated arcade is not fully ionized, though the temperature there is typically above $10^5$~K. 

To make a direct comparisons with observations, we have made synthetic images in \ion{He}{II} 304~\AA\, and H$\alpha$, which are shown in Fig.~\ref{fig2} {(e and f)}.
It is noted that neither of these two lines is usually considered optically thin.
However, for the typical environment of a quiet corona and filament, the optically thin assumption still works well in practice for \ion{He}{II} 304~\AA. This is shown in, for example, \citet{chen2015} and \citet{xia2016}.
Thus, we can simply calculate the synthetic 304~\AA\, radiation using:
\begin{equation}
I_{304}\left(x,y\right)=G_{304}\left(T\left(x,y\right)\right)n_\mathrm{e}\left(x,y\right){n_\mathrm{H}}\left(x,y\right)\Delta z\,.
  \label{eq31}
\end{equation}
The temperature-dependent response function $G_{304}(T)$ is obtained from the CHIANTI atomic database \citep{dere1997,delz2021}, where the lower temperature limit is $10^4$~K.
Therefore, for regions cooler than $10^4~\unit{K}$ in our simulations, we used the fixed value $G_{304} (T=10^4\unit{K})$ as an approximation.
In addition, a thickness of $\Delta z=10$~Mm is assumed along the LOS direction.
From Fig.~\ref{fig2}(e), we can see that the heated arcade section is slightly brighter than the background, while the prominence--corona transition region (PCTR) that envelopes the filament is the brightest part in the corona.
We note that in this 2D simulation, we cannot see the whole PCTR covering the filament, especially the part in the LOS ($z$) direction.
That explains why the filament appears as an artificially dark structure.
In fact, in a three-dimensional (3D) simulation, where we can see the whole PCTR, the filament would be a bright structure, as seen in observations.

The H$\alpha$ image in Fig.~\ref{fig2}(f) is synthesised using the approximate method in \citet{hein2015}, which is found to be a good approximation compared to a one-dimensional (1D) full radiative transfer approach \citep{jenk2022b}.
This method works well only for the filament in the corona.
Thus, synthesised radiation in the lower region, where $y<5$~Mm, may deviate from actual values.
Again, a thickness of $\Delta z=10$~Mm in the third direction is assumed.
In this panel, we can clearly see a vertical sheet-like bright structure, representing the bright prominence.

\begin{figure*}
  \centering
  \includegraphics[bb=0 60 500 1000,scale=0.7]{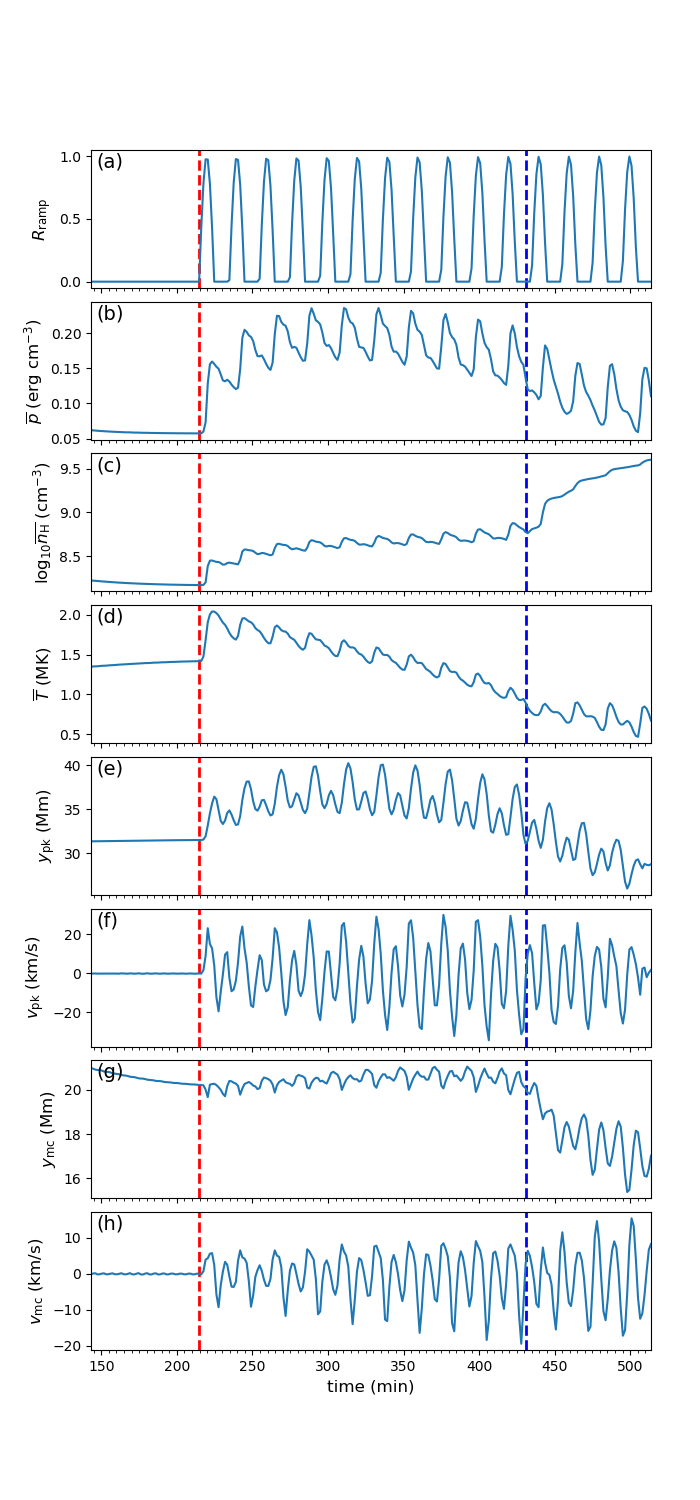}
  \caption{Time evolution of (a) $R_{\mathrm{ramp}}$, (b) average pressure, (c) average number density, (d) average temperature in the rectangle domain shown in Fig.~\ref{fig2} (a). (e)--(f) Time evolution of the $y$-position and vertical velocity of the apex of the selected magnetic field line. (g)--(h) Time evolution of the $y$-position and vertical velocity of the mass center of the selected rectangle domain. Dashed lines show the starting time of localized heating (red) and starting time of condensation (blue).}
  \label{fig3}
\end{figure*}

In order to study the formation process in detail and to better show the oscillations, we picked out a particular region, namely, the region within the black rectangle in Fig.~\ref{fig2} (a) and  we studied the property changes inside it.
Figure~\ref{fig3} {(b-d)} shows the time evolution of the average thermal pressure, $\overline{p}$, the average number density of {hydrogen,} $\overline{n_\mathrm{H}}$, and the average temperature, $\overline{T,}$ over this region, together with the time evolution of $R_{\mathrm{ramp}}$, as described in Eq.~\ref{eq212} shown in panel a.
Dashed lines in these panels show the starting point of the localized heating and the starting point of the condensation, namely, $t=214.7$~min and $t=430.8$~min, respectively.

With the evaporation proceeding, the density in the corona increases gradually{.
For the temperature and thermal pressure, these will first increase due to the evaporation, but then decrease} due to the strong radiative cooling.
Ten {periodic oscillations of various amplitudes} could be found in Fig.~\ref{fig3} (b-d), between the two dashed lines, corresponding to the parametrically imposed heating periods in panel a.

After condensation occurs, the number density increases {in steps}, with no clear period to be found (see {Fig.~\ref{fig3}, panel }c).
Meanwhile, the time evolution of thermal pressure and temperature {(panels b and d)} still show clear periods, but  they are decreasing overall.
The thermal pressure will drop to even lower than the value before we impose the localized heating.
These results are similar to our previous 1D or 2D simulations \citep{xia2011, kepp2014}.

According to the magnetic stretching mechanism mentioned in {Sect.~\ref{sec2}}, this kind of periodic heating will cause a correspondingly periodic shrinking and expansion of the magnetic field lines.
Thus, based on the MHD frozen-in theory, the coronal plasma as well as the filament will move together with the field line.
Then, we would expect to detect the filament oscillation in the vertical direction, with the same period of the periodic impulsive heating.

To show the oscillation more clearly, we first pick out the selected field line again (as described in {Sect.~\ref{sec2})}.
{T}he time evolution of the $y-$position of the apex of the selected field line, namely, $y_{\mathrm{pk}}\left(t\right)${, is shown in Fig.~\ref{fig3} (e)}.
Panel f shows the corresponding velocity $v_{\mathrm{pk}}\left(t\right)$ in the $y$-direction at this point.
The oscillation of the field line is slightly different from the physical parameters in panels (b-d).
Its oscillation period seems to be only half of the given $P$.
And every two periods are composed of a larger {peak} and a smaller one.
The oscillation of the selected field line has a typical amplitude of about 5~Mm and a velocity amplitude of about 20~\kms.
And since the plasma around this field line does not experience condensation, the oscillation pattern does not change significantly after condensation occurs on the inner field lines.

The oscillation of the filament is described by the $y$-position of the mass center $y_{\mathrm{mc}}\left(t\right)$ of the rectangle region {(see Fig.~\ref{fig2}, panel a)} as well as its corresponding density-weighted average velocity $v_{\mathrm{mc}}\left(t\right)$ in the $y$-direction.
Their time evolution could be found in Fig.~\ref{fig3} (g and h), respectively.
Similarly to the oscillation of the selected field line, the oscillation of the filament also has a period of only {1/2~}$P$.
It can be seen (especially from the $y_{\mathrm{mc}}$ curve) that the oscillation becomes stronger after condensation occurs.
The amplitude of $v_{\mathrm{mc}}$ also increases slightly, with a typical value slightly higher than 10~\kms.

\begin{figure*}
  \centering
  \includegraphics[width=16cm]{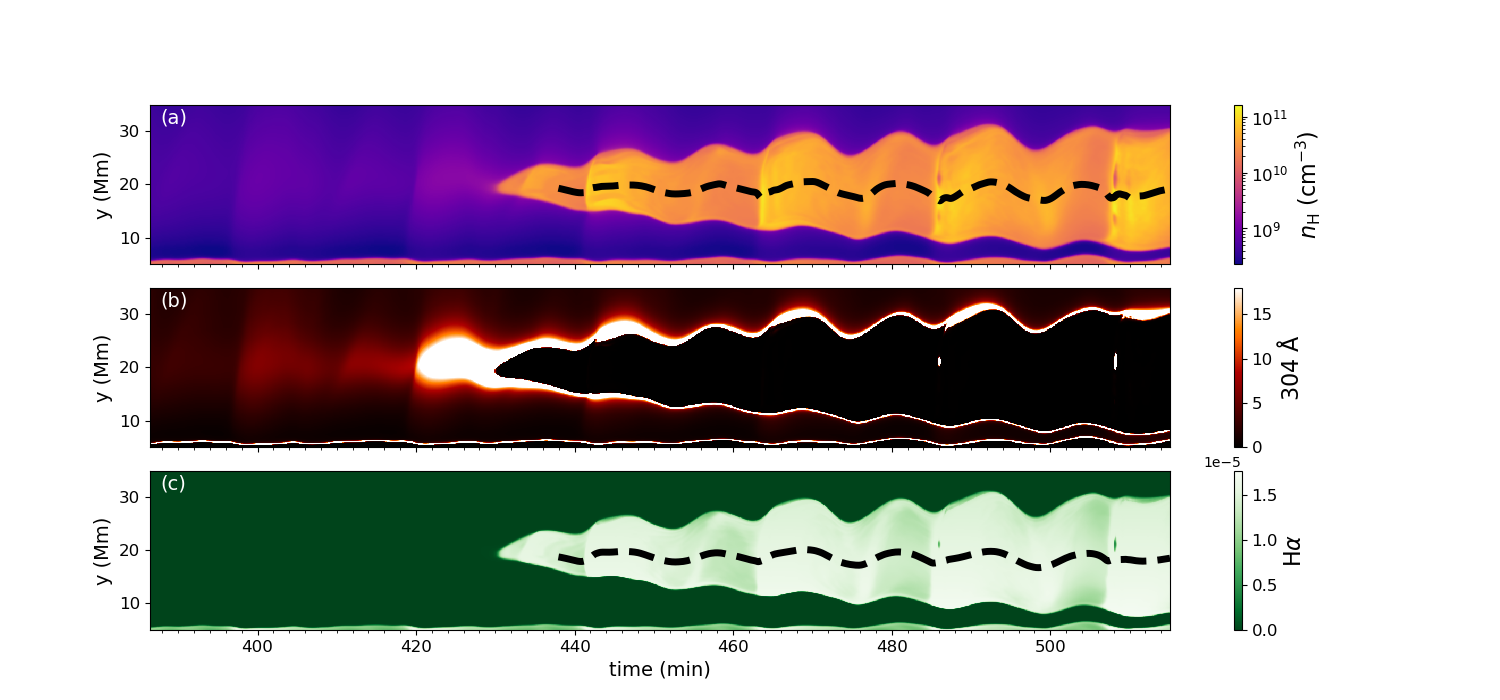}
  \caption{Time-distance plots along the $y-$axis for (a) number density, (b) 304~\AA\, synthetic radiation, and (c) H$\alpha$ synthetic radiation. Dashed lines in {panels} a and c trace the density-weighted and radiation-weighted center position, respectively, after the condensation.}
  \label{fig4}
\end{figure*}

To demonstrate the oscillation of the filament more clearly, we took a {vertical} slice along {$x = 0$}, and made a time-distance plot in Fig.~\ref{fig4}.
The evolution of the number density of the filament, synthesised 304~\AA\, radiation and H$\alpha$ radiation along this slice are shown in panels a-c, respectively.
From these panels, we can clearly see the oscillation during the formation and growth of the filament.
In the number density panel, we also drew the $y-$position of the mass center using dashed lines.
In this way, we can clearly see that the period of the oscillation is not $P$, but approximately {1/2~}$P$.
A similar dashed line, which is weighted by the H$\alpha$ radiation, is shown in panel c, displaying a similar oscillation.
In the case of the 304~\AA\, panel, the bright PCTR structure shows the oscillation with period of approximately {1/2~}$P$, similarly to the density and H$\alpha$ plots.

\begin{figure*}
  \centering
  \includegraphics[width=16cm]{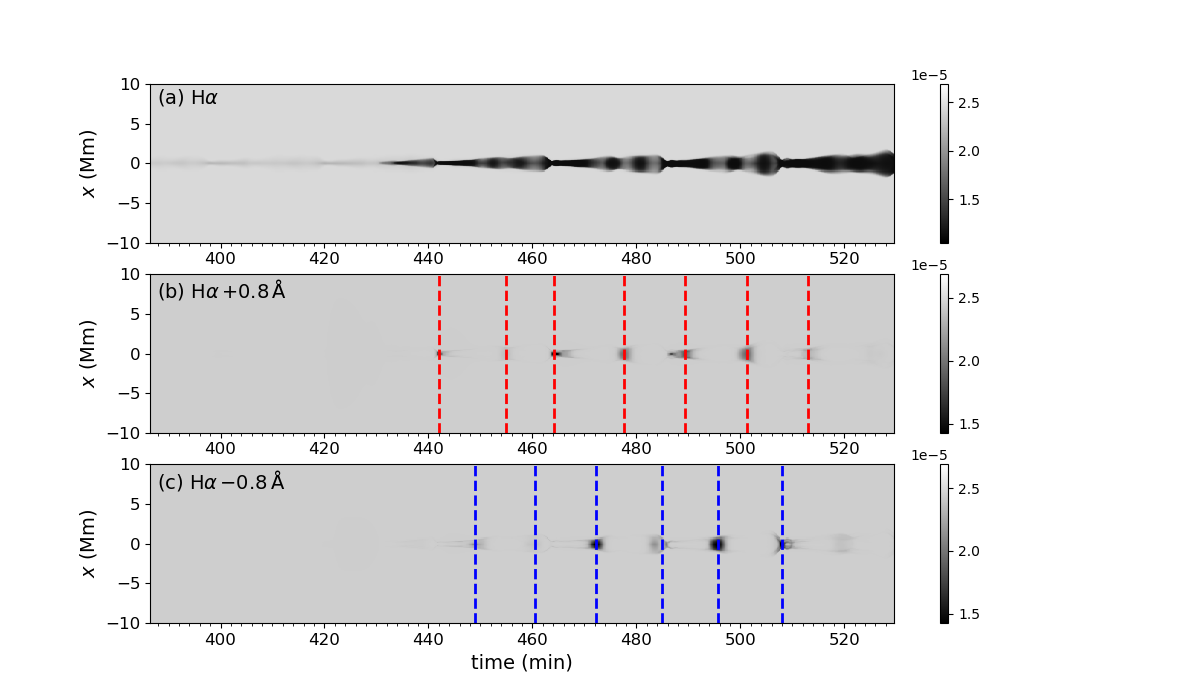}
  \caption{Time evolution of the synthetic (a) H$\alpha$ line center (b) H$\alpha$ red wing (c) H$\alpha$ blue wing radiation from the filament view (top-down view). Vertical dashed lines are drawn to help to see the periods.}
  \label{fig5}
\end{figure*}

We refer to the filament oscillating in the vertical direction as ``winking'' because it will appear and disappear periodically in the H$\alpha$ line center and wings (especially).
To confirm this, in addition to the oscillation in the prominence view (face-on view), we must check the filament view (top-down view).
The synthetic H$\alpha$ radiation in the filament view is also calculated following the method in \citet{hein2015}, for both H$\alpha$ line center and {wings at wavelengths $\pm0.8$\ \AA.
Observations of prominences and filaments by  \citet{eto2002} and \citet{okam2004} were also made in these wavelengths.}
However, in that view, the chromosphere (the region below {$y = 2.5$~Mm}) is not included in this upward integration since the emission from the lower atmosphere is treated as background emission.
Our simulation is a 2D simulation, and when observed from the top, the data cube will collapse into an integrated emission sheet.
We show the time evolution of this H$\alpha$ sheet in Fig.~\ref{fig5} (a).

Before condensation occurs, nothing special could be observed.
After the condensation, we can see a "black cloud" appear in the figure.
The area of the H$\alpha$ filament changes periodically, becoming sometimes small and faint while sometimes large and clear.
The maximum area is typically 3-4 times bigger than the minimum area.
Typically, in this case, the filament always exists in the H$\alpha$ band.
In some of our tests (see {Sect.~\ref{sec4}}), it will be very faint and nearly invisible.
In observations, both situations could occur (see also the references listed in Table~\ref{tb1}).

Synthetic filament views observed in the red and blue wings are presented in Fig.~\ref{fig5} (b and c), respectively.
Similarly to the previous oscillation patterns, the period here is typically {1/2~}$P$.
It is clear in these two panels that the filament is visible in the red wing and blue wing alternately, which is consistent with the observations.
Thus, the simulated filament can be called a winking filament.


\section{Discussion and conclusion}
\label{sec4}
\subsection{Details of the magnetic stretching}
{In this part of the discussion, we look to find an answer to the question of why magnetic field lines become stretched during the evaporation.}
The physical scenario should be as follows.
When the lower atmosphere is heated by the localized heating, the local pressure will increase so that it can produce a pressure gradient, by which plasma will be pushed to move upward, generating upward flow.
This is the so-called evaporation.
{T}his pressure gradient is typically directed {vertically upwards (i.e., along the positive $y-$direction).
However, the magnetic field lines are inclined, having an angle ($x-$varying)} with the vertical $y-$direction.
According to the MHD frozen-in theorem, the upward moving plasma will rise together with the magnetic field lines, dragging it into a more stretched state. We can illustrate the analogy with an inflated balloon: as pressure rises below, the line-tied magnetic field lines stretch outwards. We go  on to analyze this in detail for a simple, but representative, ramped-up heating phase.

\begin{figure*}
  \centering
  \includegraphics[bb=0 60 600 1000,scale=0.7]{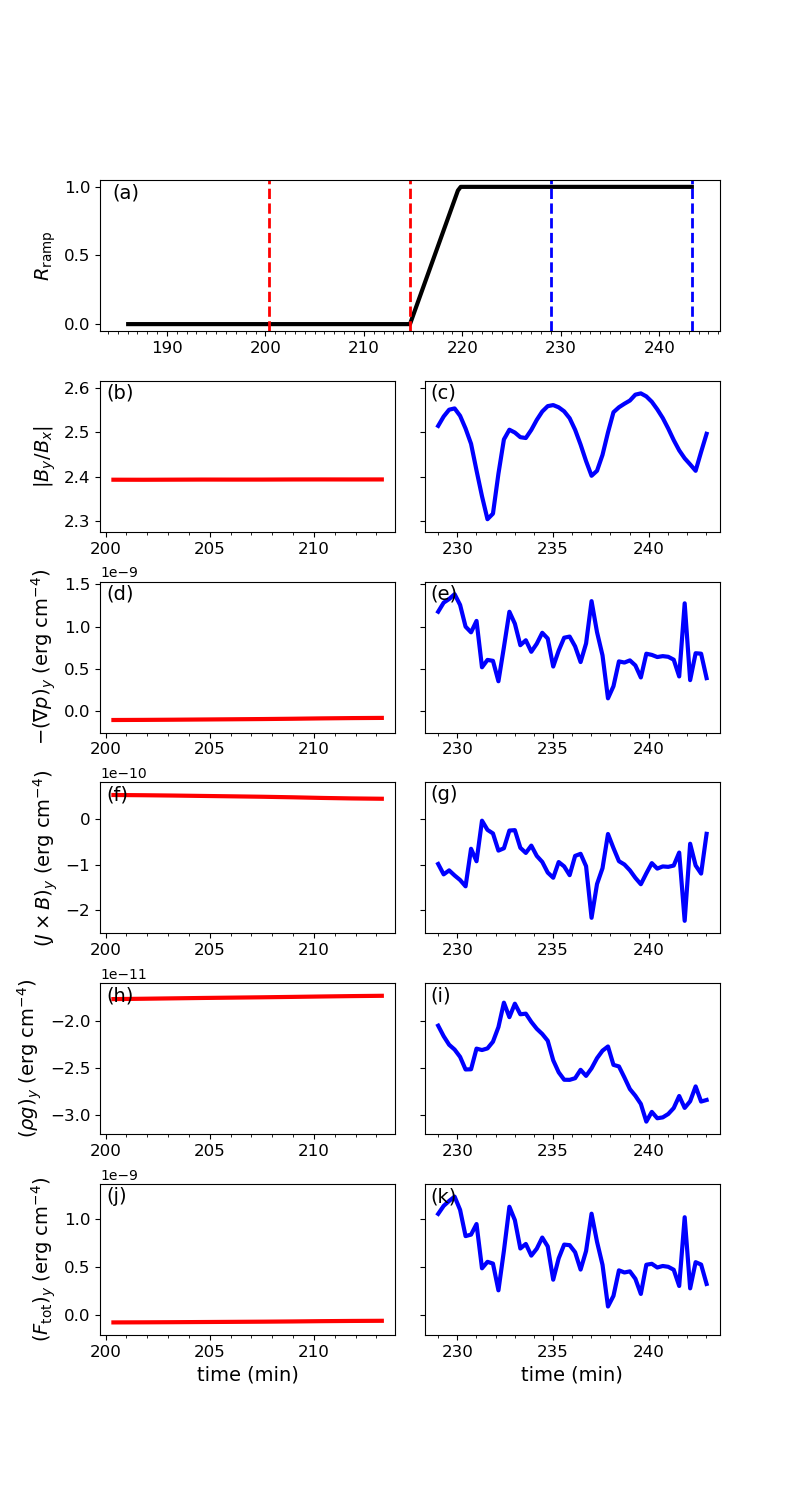}
  \caption{Time evolution of $R_{\mathrm{ramp}}$ shown in panel a. Time evolution of (b) $\left| {{B_y}/{B_x}} \right|$, (d) gas pressure gradient in the $y-$direction, (f) Lorentz force in the $y-$direction, (h) gravity force, (j) total force in the $y-$direction before the localized heating is imposed are shown on the left column. Panels in the right column show the time evolution of these parameters after the localized heating is on, correspondingly.}
  \label{fig6}
\end{figure*}

Therefore, instead of the periodic $R_{\mathrm{ramp}}$ as used in the previous sections,  {here we} use a linear $R_{\mathrm{ramp}}$ function instead, whereby:

\begin{equation}
R_{\mathrm{ramp}}\left( t \right) = 
\left\{ 
{
\begin{array}{*{20}{c}}
4\left( {t - {t_{\mathrm{relax}}}} \right)/P & \text{if } t_{\mathrm{relax}} < t < t_{\mathrm{relax}}+P/4,\\
1 & \text{if } t \geq t_{\mathrm{relax}}+P/4 \,.
\end{array}
} 
\right.
\label{eq41}
\end{equation}
This new $R_{\mathrm{ramp}}(t)$ function is shown in Fig.~\ref{fig6} (a), and grows from 0 to 1 linearly in 5~min, that is, during $t=214.7$~min and $t=219.7$~min.
We again selected the field line starting from $(x, y) = (45, 0)$~Mm and analyzed the forces at a particular height on this line, for instance, at $y=4$~Mm on the right side.
During the simulation, the magnetic field line deforms and stretches, so that its position moves in the horizontal direction.
Therefore, it cannot always be the "same" point.
Considering that the deformation of the field line at the lower atmosphere is minor, we treated the selected point as the same point approximately, as indicated by the blue circle in Fig.~\ref{fig1} (d).

In the second row, we show the time evolution of the inclination of the magnetic field line at this point in the two steady heating phases, namely, before (during relaxation) and after the ramping up of the localized heating.
In panel b, before we introduce the localized heating, namely, prior to $t=214.7$~min, $\left| {{B_y}/{B_x}} \right|$, which indicates the inclination of the magnetic field line at the selected point, is typically 2.4.
This increases a little bit after the heating is fully on (panel c), which means that the magnetic field line becomes more vertical.
Hence, it gets stretched in the $y-$direction.
In panel d the $y-$component of pressure gradient, namely, $- \left( {\nabla p} \right)_{y}$, is typically $-3\times 10^{-10}$ in cgs units at the end of the relaxation phase, but increases to about $5\times10^{-10}$ (panel e), showing a significant increase.
Although it is not shown here, the $x-$component of the pressure gradient shows the same tendency.
That means during the relaxation phase, the pressure gradient is slightly inwards (negative $- \left( {\nabla p} \right)_{x,y}$ values) for the selected loop, but its direction turns outward in the evaporation phase.
Correspondingly, the Lorentz force $\bma{J} \times \bma{B}$ changes in the opposite direction to ensure net overall force balance, from slightly directing outward to pointing inward (panels f and g).
Meanwhile, the evolution of the gravity force $\left(\rho \bma{g}\right)_y$ shows a similar response to the localized heating as the Lorentz force ({panels} h and i).
This means that before the localized heating is imposed, {the gravity together with the gas pressure gradient exerts a force to collapse the loop.
Meanwhile,} the Lorentz force plays a supporting role.
In panel j, we can clearly see that the total force $F_{\mathrm{tot}}$ {(in the $y-$direction)} is nearly balanced during the relaxation.
Then, after the extra localized heating is on, the gas pressure gradient pushes the plasma moving upward, trying to expand the loop, while the Lorentz force will suppress such a motion, and changes its direction to act downwards.
Such a clear change in the interaction of all forces eventually causes the change in the shape (i.e., the local direction) of the magnetic field.
We note that the evolution of the total force (panel k) is very close to the evolution of pressure gradient (panel e), indicating the most important role played by the pressure gradient here.

\subsection{Parameter survey}
Since the stretching of the magnetic field line is triggered by the evaporation flow caused by the upward pressure gradient, plasma $\beta$ will definitely be important in this physical process.
To see how different plasma $\beta$s would influence the result of our simulation, we can choose to use different magnetic field strength, {namely,} $B_0$, or alternatively, to put the localized heating at different positions.
These two methods are not totally equivalent, but both of them should work.
Here, for convenience, we chose to put the localized heating at different positions, to simulate heating at different heights.
For all these different runs, we simply compare the heights of the apex of the selected field line.
The results are shown in Table~\ref{tb2}.


\begin{table*}[h]
\newcommand{\tablecell}[2]{\begin{tabular}{@{}#1@{}}#2\end{tabular}}
\centering
\caption{Heating at different positions.}
\label{tb2}
\begin{tabular}{c|c|c|c|c|c|c}
\hline
run & \tablecell{c}{$x_r$ \\ (Mm)} & \tablecell{c}{$y_1$ \\ (Mm)} & plasma $\beta$ & \tablecell{c}{$v_y$($y=10$~Mm) \\ (\kms)}  & \tablecell{c}{$y_{\mathrm{pk}}$($t$=214.7~min) \\ (Mm)} & \tablecell{c}{$y_{\mathrm{pk}}$($t$=243.3~min)  \\ (Mm)}\\
\hline
A1 & 42.54 & 2.0 & 0.1188 & 37.60 & 31.48 & 34.80 \\
A2 & 42.29 & 2.5 & 0.1359 & 40.48 & 31.48 & 35.91 \\
A3 & 42.04 & 3.0 & 0.1399 & 42.49 & 31.48 & 37.20 \\
A4 & 41.78 & 3.5 & 0.1422 & 50.41 & 31.48 & 38.60 \\
A5 & 41.50 & 4.0 & 0.1451 & 58.43 & 31.48 & 40.06 \\
A6 & 41.23 & 4.5 & 0.1482 & 60.05 & 31.48 & 41.55 \\
A7 & 40.94 & 5.0 & 0.1514 & 66.31 & 31.48 & 43.04 \\
A8 & 40.65 & 5.5 & 0.1547 & 66.88 & 31.48 & 44.62 \\
A9 & 40.35 & 6.0 & 0.1581 & 69.11 & 31.48 & 46.16 \\
\hline
\end{tabular}
\end{table*}

The localized heating typically takes place at the chromosphere and TR.
Thus, we chose for $y_1$ in Eq.~\ref{eq210} {to change} from 2~Mm to 6~Mm, while the $x-$position (e.g., $x_r$) is changed correspondingly along the field line ($x_l=-x_r$).
The corresponding plasma $\beta$ at ($x_r$, $y_1$) ranges from 0.1188 to 0.1581, {thus, it changes} only slightly.
However, we can see that in different runs, the vertical velocity $v_y$ at $y=10$~Mm (on the same field line connecting $(x_r, y_1)$) increases from 37.6~\kms to 69.11~\kms.
In addition, the apex of the selected field line {$y_{\mathrm{\mathrm{pk}}}(t=243.3$~min)} also rises significantly.
This means that when the same magnitude of localized heating is acting at a lower position in the atmosphere, the stretching effect becomes progressively minor.
When it takes place in the higher chromosphere or TR, the stretching can work very effectively.

Besides plasma $\beta$, another important ingredient of this mechanism is the strength of the heating, $H_1$.
It represents the energy release rate throughout the lower atmosphere.
The results of different runs with different $H_1$ are listed in Table~\ref{tb3}, similar with the values used in previous works \citep{xia2011,pelo2022}.
For localized heating injected into the system at the same location, stronger heating will generate higher upflow velocity and the magnetic field lines will get more stretched.
However, while stronger heating will induce a more effective magnetic stretching, it will prohibit condensation from occurring, as demonstrated by \citet{xia2011}.

Notice should be taken that, the test runs here quantify the oscillation of field lines before the formation of filaments.
Therefore, these runs cannot yet be compared directly to the observations, but if we were to continue these runs using similar periodic cycling, the same winking effects on the filaments (once formed) can be expected.


\begin{table*}[h]
\newcommand{\tablecell}[2]{\begin{tabular}{@{}#1@{}}#2\end{tabular}}
\centering
\caption{Different heating strengths.}
\label{tb3}
\begin{tabular}{c|c|c|c|c}
\hline
run & \tablecell{c}{$H_1$ \\ (erg s$^{-1}$ cm$^{-3}$)}  & \tablecell{c}{$v_y$($y=10$~Mm) \\(\kms)}  & \tablecell{c}{$y_{\mathrm{pk}}$($t$=214.7~min) \\(Mm)} & \tablecell{c}{$y_{\mathrm{pk}}$($t$=243.3~min) \\(Mm)} \\
\hline
B1 & 0.0013 & 6.31   &  31.48 & 32.10 \\
B2 & 0.0025 & 9.75   &  31.48 & 32.67 \\
B3 & 0.0050 & 16.80  &  31.48 & 33.68 \\
B4 & 0.0100 & 29.06  &  31.48 & 35.74 \\
B5 & 0.0200 & 58.43  &  31.48 & 40.06 \\
B6 & 0.0400 & 75.35  &  31.48 & 47.35 \\
B7 & 0.0800 & 107.80 &  31.48 & 55.56 \\
B8 & 0.1600 & 126.05 &  31.48 & 61.58 \\
B9 & 0.3200 & 104.53 &  31.48 & 68.18 \\
\hline
\end{tabular}
\end{table*}

In this work, we focus on the evaporation\discretionary{-}{-}{-}condensation mechanism.
A similar mechanism for filament formation is the injection model, in which thermal instability is not necessary, and energetic events in the lower atmosphere will push directly the cool material into the solar corona.
From the ``unified model'' for prominence formation \citep{huan2021}, the difference between the evaporation\discretionary{-}{-}{-}condensation model and the injection model should typically be achieved by varying the values of $y_1$ and $H_1$ here.
The injection model usually adopts a lower $y_1$ and stronger $H_1$.
Thus, in principle, the injection model could also produce such kind of vertical oscillations.

\subsection{Driven period}
The period of the impulsive heating $P$ is chosen to be 20~min in this work.
However, the period of the winking filament seems to be approximately only {half of this value} -- and this is not a surprise.
We should distinguish three types of periods, or frequencies here: the frequency of the impulsive heating or driven force, $\omega_{\mathrm{dirv}}$, the frequency, $\omega_{\mathrm{wink}}$, of the observed oscillation of the winking filament, and the eigenfrequency, $\omega_0$, of the magnetic loop.
In the research of coronal seismology, especially the kink mode, $\omega_{\mathrm{dirv}}$ is usually equal to $\omega_0$ \citep[see recent reviews in e.g.,][]{vand2020, naka2021}. This is because the loop system will naturally select out a narrow band frequency close to the eigenfrequency, so that the kink mode can work efficiently.

However, in our simulation, the frequency of the additional heating $\omega_{\mathrm{dirv}}$ is more likely to be inconsistent with $\omega_0$.
A similar study was recently carried out by \citet{ni2022} where the authors use periodic jets to drive the oscillation of filaments. In a more general sense, this scenario is similar to the {forced} oscillators in the classic mechanics, where the solution usually takes the following form:

\begin{equation}
\begin{aligned}
  s(t) &= \frac{f}{{\omega _0^2 - {\omega _{\mathrm{dirv}}}^2}}\left( {\cos {\omega _{\mathrm{dirv}}}t - \cos {\omega _0}t} \right), \\ &= \frac{{2f}}{{\omega _0^2 - {\omega _{\mathrm{dirv}}}^2}}\sin \frac{{\left( {{\omega _0} - {\omega _{\mathrm{dirv}}}} \right)t}}{2}\sin \frac{{\left( {{\omega _0} + {\omega _{\mathrm{dirv}}}} \right)t}}{2} \,,
  \label{eq42}
\end{aligned}
\end{equation}
where $s(t)$ is the displacement of the oscillator and $f$ is the driven force per mass.
According to Eq.~\ref{eq42}, the observed frequency $\omega_{\mathrm{wink}}$ is composed of two frequencies: ${\left( {{\omega _0} - {\omega _{\mathrm{dirv}}}} \right)}/2$ and ${\left( {{\omega _0} + {\omega _{\mathrm{dirv}}}} \right)}/2$.
We note that the amplitude of the oscillation is inversely proportional to ${\left( {{\omega _0^2} - {\omega _{\mathrm{dirv}}}^2} \right)}$.
Therefore, $\omega _{\mathrm{dirv}}$ cannot deviate from $\omega _{0}$ too much.
With this assumption, then, considering that we would usually recognize the higher frequency in observations, we have:
\begin{equation}
{\omega _{\mathrm{wink}}} = \left( {{\omega _0} + {\omega _{\mathrm{dirv}}}} \right)/2 \,.
\label{eq43}
\end{equation}
We can take $\omega _0=2\pi/P_0=v_A\pi/L$, which is usually used in the study of coronal seismology \citep{robe1984}, as an approximation. Then, $P_0$ is the corresponding eigenperiod, $v_A$ is the Alfvén speed, and $L$ is the length of the loop.
In our simulations, $\omega_{\mathrm{dirv}}$ is fixed while $\omega _0$ could be different from field line to field line and from time to time.
However, before the formation of filament (i.e., between $t=224.7$~min and $t=430.8$~min), the eigenperiod $P_0$ within the region where our filament later forms{} is typically around 450 s and gradually drops to around 330 s after  condensation.
Using Eq.~\ref{eq42}, we then {get the} observed period $P_{\mathrm{wink}}$ {of the winking from the range} 542~s to 655~s, which is approximately 10 min or {1/2~}$P$. This value is similar {to} our results.

Therefore, it is also not necessary to have the period of the impulsive heating {of} exactly 20~min, and even the impulsive heating does not necessarily need to be periodic.
A quasi-periodic heating should be enough to get the winking phenomenon.

\subsection{Conclusion}

Within the one-fluid ideal MHD description, magnetic field lines move together with the plasma.
Small energy events in the lower atmosphere can produce localized heating that increases the local pressure gradient, causing the evaporation of the plasma.
Since the magnetic field is coupled with the plasma, the plasma can move freely along the field line, and at the same time, the field line will be dragged by any upward moving plasma. We have shown that at least a 2D (but completely analogous in 3D) model is required to show that line-tied stretching {of field line} is inevitable whenever a localized heating turns on, beyond the background coronal heating.
This process has been overlooked by all restricted 1D models, which reduce coronal dynamics and prominence formation scenarios in strong field settings to 1D hydrodynamic along a fixed, prechosen field line shape.

In this work, we demonstrate that the evaporation\discretionary{-}{-}{-}condensation scenario of prominence formation, during which a small portion of energy is injected into the low atmosphere, can produce oscillations of the filament in the vertical direction.
When the heating source then acts (quasi-) periodically, the winking filament phenomenon can be observed.
However, this also depends on the position and strength of the heating source.

In our simulation, we found that the periods of impulsive heating and the oscillation{s} of the filament are {equal to each other, and period of the oscillations is} typically reduce{d} by a factor of two.
This must relate to the eigenfrequency of the perturbed loop being different from the frequency of the impulsive heating.
The interplay and beating of two frequencies result in the altered period of oscillations that we see in our simulations.


In previous observations, a winking filament is usually interpreted as being caused by waves or energetic events.
Here, we demonstrate through numerical simulations that winking filaments may also be observed due to the evaporation during the formation phase of filaments.
As far as we know, this has not yet been reported in observations, {but we expect} that future observations will be able to fully confirm {} this scenario.

\begin{acknowledgements}
        
      We thank A. Hillier, S. Gunár, M. Guo for valuable suggestions.
      {We thank the referee for detailed suggestions on writing standard.}
      YZ acknowledges funding from Research Foundation – Flanders FWO under the project number 1256423N.
      XL and RK acknowledge the European Research Council (ERC) under the European Unions Horizon 2020 research and innovation program (grant agreement No. 833251 PROMINENT ERC-ADG 2018).
      JH acknowledges funding from NSFC under grant 11903020.
      RK acknowledges support by Internal funds KU Leuven, project C14/19/089 TRACESpace and FWO project G0B4521N. 
      Visualisations used the open source software \href{https://www.python.org/}{Python}. Resources and services used in this work were provided by the VSC (Flemish Supercomputer Center), funded by the Research Foundation - Flanders (FWO) and the Flemish Government.
\end{acknowledgements}

\bibliographystyle{aa}
\bibliography{ref}

\begin{thebibliography}{75}
\expandafter\ifx\csname natexlab\endcsname\relax\def\natexlab#1{#1}\fi

\bibitem[{{Amari} {et~al.}(2014){Amari}, {Canou}, \& {Aly}}]{amar2014}
{Amari}, T., {Canou}, A., \& {Aly}, J.-J. 2014, \nat, 514, 465

\bibitem[{{An} {et~al.}(1988){An}, {Bao}, \& {Wu}}]{an1988}
{An}, C.-H., {Bao}, J.~J., \& {Wu}, S.~T. 1988, \solphys, 115, 81

\bibitem[{{Antiochos} {et~al.}(2000){Antiochos}, {MacNeice}, \&
  {Spicer}}]{anti2000}
{Antiochos}, S.~K., {MacNeice}, P.~J., \& {Spicer}, D.~S. 2000, \apj, 536, 494

\bibitem[{{Antiochos} {et~al.}(1999){Antiochos}, {MacNeice}, {Spicer}, \&
  {Klimchuk}}]{anti1999}
{Antiochos}, S.~K., {MacNeice}, P.~J., {Spicer}, D.~S., \& {Klimchuk}, J.~A.
  1999, \apj, 512, 985

\bibitem[{{Arregui} {et~al.}(2018){Arregui}, {Oliver}, \&
  {Ballester}}]{arre2018}
{Arregui}, I., {Oliver}, R., \& {Ballester}, J.~L. 2018, Living Reviews in
  Solar Physics, 15, 3

\bibitem[{{Asai} {et~al.}(2012){Asai}, {Ishii}, {Isobe}, {Kitai}, {Ichimoto},
  {UeNo}, {Nagata}, {Morita}, {Nishida}, {Shiota}, {Oi}, {Akioka}, \&
  {Shibata}}]{asai2012}
{Asai}, A., {Ishii}, T.~T., {Isobe}, H., {et~al.} 2012, \apjl, 745, L18

\bibitem[{{Aschwanden}(2001)}]{asch2001}
{Aschwanden}, M.~J. 2001, \apj, 560, 1035

\bibitem[{{Ballester}(2005)}]{ball2005}
{Ballester}, J.~L. 2005, \ssr, 121, 105

\bibitem[{{Berger} {et~al.}(2017){Berger}, {Hillier}, \& {Liu}}]{berg2017}
{Berger}, T., {Hillier}, A., \& {Liu}, W. 2017, \apj, 850, 60

\bibitem[{{Berger} {et~al.}(2011){Berger}, {Testa}, {Hillier}, {Boerner},
  {Low}, {Shibata}, {Schrijver}, {Tarbell}, \& {Title}}]{berg2011}
{Berger}, T., {Testa}, P., {Hillier}, A., {et~al.} 2011, \nat, 472, 197

\bibitem[{{Berger} {et~al.}(2010){Berger}, {Slater}, {Hurlburt}, {Shine},
  {Tarbell}, {Title}, {Lites}, {Okamoto}, {Ichimoto}, {Katsukawa}, {Magara},
  {Suematsu}, \& {Shimizu}}]{berg2010}
{Berger}, T.~E., {Slater}, G., {Hurlburt}, N., {et~al.} 2010, \apj, 716, 1288

\bibitem[{{Berghmans} {et~al.}(2021){Berghmans}, {Auch{\`e}re}, {Long},
  {Soubri{\'e}}, {Mierla}, {Zhukov}, {Sch{\"u}hle}, {Antolin}, {Harra},
  {Parenti}, {Podladchikova}, {Aznar Cuadrado}, {Buchlin}, {Dolla}, {Verbeeck},
  {Gissot}, {Teriaca}, {Haberreiter}, {Katsiyannis}, {Rodriguez}, {Kraaikamp},
  {Smith}, {Stegen}, {Rochus}, {Halain}, {Jacques}, {Thompson}, \&
  {Inhester}}]{berg2021}
{Berghmans}, D., {Auch{\`e}re}, F., {Long}, D.~M., {et~al.} 2021, \aap, 656, L4

\bibitem[{{Brughmans} {et~al.}(2022){Brughmans}, {Jenkins}, \&
  {Keppens}}]{brug2022}
{Brughmans}, N., {Jenkins}, J.~M., \& {Keppens}, R. 2022, \aap, 668, A47

\bibitem[{{Carlsson} \& {Leenaarts}(2012)}]{carl2012}
{Carlsson}, M. \& {Leenaarts}, J. 2012, \aap, 539, A39

\bibitem[{{Chae}(2003)}]{chae2003}
{Chae}, J. 2003, \apj, 584, 1084

\bibitem[{{Chen} {et~al.}(2015){Chen}, {Peter}, {Bingert}, \&
  {Cheung}}]{chen2015}
{Chen}, F., {Peter}, H., {Bingert}, S., \& {Cheung}, M.~C.~M. 2015, Nature
  Physics, 11, 492

\bibitem[{{Chen} {et~al.}(2002){Chen}, {Wu}, {Shibata}, \& {Fang}}]{chen2002}
{Chen}, P.~F., {Wu}, S.~T., {Shibata}, K., \& {Fang}, C. 2002, \apjl, 572, L99

\bibitem[{{Chen} {et~al.}(2020){Chen}, {Xu}, \& {Ding}}]{chen2020}
{Chen}, P.-F., {Xu}, A.-A., \& {Ding}, M.-D. 2020, Research in Astronomy and
  Astrophysics, 20, 166

\bibitem[{{Colgan} {et~al.}(2008){Colgan}, {Abdallah}, {Sherrill}, {Foster},
  {Fontes}, \& {Feldman}}]{colg2008}
{Colgan}, J., {Abdallah}, J., J., {Sherrill}, M.~E., {et~al.} 2008, \apj, 689,
  585

\bibitem[{{Dalgarno} \& {McCray}(1972)}]{dalg1972}
{Dalgarno}, A. \& {McCray}, R.~A. 1972, \araa, 10, 375

\bibitem[{{De Groof} \& {Goossens}(2002)}]{degr2002}
{De Groof}, A. \& {Goossens}, M. 2002, \aap, 386, 691

\bibitem[{{Del Zanna} {et~al.}(2021){Del Zanna}, {Dere}, {Young}, \&
  {Landi}}]{delz2021}
{Del Zanna}, G., {Dere}, K.~P., {Young}, P.~R., \& {Landi}, E. 2021, \apj, 909,
  38

\bibitem[{{Dere} {et~al.}(1997){Dere}, {Landi}, {Mason}, {Monsignori Fossi}, \&
  {Young}}]{dere1997}
{Dere}, K.~P., {Landi}, E., {Mason}, H.~E., {Monsignori Fossi}, B.~C., \&
  {Young}, P.~R. 1997, \aaps, 125, 149

\bibitem[{Dyson(1930)}]{dyso1930}
Dyson, F. 1930, MN, 91, 239

\bibitem[{{Eto} {et~al.}(2002){Eto}, {Isobe}, {Narukage}, {Asai}, {Morimoto},
  {Thompson}, {Yashiro}, {Wang}, {Kitai}, {Kurokawa}, \& {Shibata}}]{eto2002}
{Eto}, S., {Isobe}, H., {Narukage}, N., {et~al.} 2002, \pasj, 54, 481

\bibitem[{{Fan}(2018)}]{fan2018}
{Fan}, Y. 2018, \apj, 862, 54

\bibitem[{{Gilbert} {et~al.}(2008){Gilbert}, {Daou}, {Young}, {Tripathi}, \&
  {Alexander}}]{gilb2008}
{Gilbert}, H.~R., {Daou}, A.~G., {Young}, D., {Tripathi}, D., \& {Alexander},
  D. 2008, \apj, 685, 629

\bibitem[{{Grechnev} {et~al.}(2014){Grechnev}, {Uralov}, {Chertok}, {Slemzin},
  {Filippov}, {Egorov}, {Fainshtein}, {Afanasyev}, {Prestage}, \&
  {Temmer}}]{grec2014}
{Grechnev}, V.~V., {Uralov}, A.~M., {Chertok}, I.~M., {et~al.} 2014, \solphys,
  289, 1279

\bibitem[{{Heinzel} {et~al.}(2015){Heinzel}, {Gun{\'a}r}, \&
  {Anzer}}]{hein2015}
{Heinzel}, P., {Gun{\'a}r}, S., \& {Anzer}, U. 2015, \aap, 579, A16

\bibitem[{{Hong} {et~al.}(2022){Hong}, {Carlsson}, \& {Ding}}]{hong2022}
{Hong}, J., {Carlsson}, M., \& {Ding}, M.~D. 2022, \aap, 661, A77

\bibitem[{{Howson} {et~al.}(2019){Howson}, {De Moortel}, {Antolin}, {Van
  Doorsselaere}, \& {Wright}}]{hows2019}
{Howson}, T.~A., {De Moortel}, I., {Antolin}, P., {Van Doorsselaere}, T., \&
  {Wright}, A.~N. 2019, \aap, 631, A105

\bibitem[{{Huang} {et~al.}(2021){Huang}, {Guo}, {Ni}, {Xu}, \&
  {Chen}}]{huan2021}
{Huang}, C.~J., {Guo}, J.~H., {Ni}, Y.~W., {Xu}, A.~A., \& {Chen}, P.~F. 2021,
  \apjl, 913, L8

\bibitem[{{Hyder}(1966)}]{hyde1966}
{Hyder}, C.~L. 1966, \zap, 63, 78

\bibitem[{{Isobe} \& {Tripathi}(2006)}]{isob2006}
{Isobe}, H. \& {Tripathi}, D. 2006, \aap, 449, L17

\bibitem[{{Jackiewicz} \& {Balasubramaniam}(2013)}]{jack2013}
{Jackiewicz}, J. \& {Balasubramaniam}, K.~S. 2013, \apj, 765, 15

\bibitem[{{Jenkins} \& {Keppens}(2022)}]{jenk2022a}
{Jenkins}, J.~M. \& {Keppens}, R. 2022, Nature Astronomy, 6, 942

\bibitem[{{Jenkins} {et~al.}(2023){Jenkins}, {Osborne}, \&
  {Keppens}}]{jenk2022b}
{Jenkins}, J.~M., {Osborne}, C.~M.~J., \& {Keppens}, R. 2023, \aap, 670, A179

\bibitem[{{Jer{\v{c}}i{\'c}} \& {Keppens}(2023)}]{jerv2022}
{Jer{\v{c}}i{\'c}}, V. \& {Keppens}, R. 2023, \aap, 670, A64

\bibitem[{{Jiang} {et~al.}(2021){Jiang}, {Feng}, {Liu}, {Yan}, {Hu}, {Moore},
  {Duan}, {Cui}, {Zuo}, {Wang}, \& {Wei}}]{jian2021}
{Jiang}, C., {Feng}, X., {Liu}, R., {et~al.} 2021, Nature Astronomy, 5, 1126

\bibitem[{{Johnston} {et~al.}(2019){Johnston}, {Cargill}, {Antolin}, {Hood},
  {De Moortel}, \& {Bradshaw}}]{john2019}
{Johnston}, C.~D., {Cargill}, P.~J., {Antolin}, P., {et~al.} 2019, \aap, 625,
  A149

\bibitem[{{Kaneko} \& {Yokoyama}(2015)}]{kane2015}
{Kaneko}, T. \& {Yokoyama}, T. 2015, \apj, 806, 115

\bibitem[{{Keppens} \& {Xia}(2014)}]{kepp2014}
{Keppens}, R. \& {Xia}, C. 2014, \apj, 789, 22

\bibitem[{{Klimchuk} {et~al.}(2010){Klimchuk}, {Karpen}, \&
  {Antiochos}}]{klim2010}
{Klimchuk}, J.~A., {Karpen}, J.~T., \& {Antiochos}, S.~K. 2010, \apj, 714, 1239

\bibitem[{{Li} {et~al.}(2022){Li}, {Keppens}, \& {Zhou}}]{li2022}
{Li}, X., {Keppens}, R., \& {Zhou}, Y. 2022, \apj, 926, 216

\bibitem[{{Mackay} {et~al.}(2010){Mackay}, {Karpen}, {Ballester}, {Schmieder},
  \& {Aulanier}}]{mack2010}
{Mackay}, D.~H., {Karpen}, J.~T., {Ballester}, J.~L., {Schmieder}, B., \&
  {Aulanier}, G. 2010, \ssr, 151, 333

\bibitem[{{Mandrini} {et~al.}(2000){Mandrini}, {D{\'e}moulin}, \&
  {Klimchuk}}]{mand2000}
{Mandrini}, C.~H., {D{\'e}moulin}, P., \& {Klimchuk}, J.~A. 2000, \apj, 530,
  999

\bibitem[{{Morimoto} \& {Kurokawa}(2003)}]{mori2003}
{Morimoto}, T. \& {Kurokawa}, H. 2003, \pasj, 55, 503

\bibitem[{{M{\"u}ller} {et~al.}(2004){M{\"u}ller}, {Peter}, \&
  {Hansteen}}]{mull2004}
{M{\"u}ller}, D.~A.~N., {Peter}, H., \& {Hansteen}, V.~H. 2004, \aap, 424, 289

\bibitem[{{Nakariakov} {et~al.}(2021){Nakariakov}, {Anfinogentov}, {Antolin},
  {Jain}, {Kolotkov}, {Kupriyanova}, {Li}, {Magyar}, {Nistic{\`o}}, {Pascoe},
  {Srivastava}, {Terradas}, {Vasheghani Farahani}, {Verth}, {Yuan}, \&
  {Zimovets}}]{naka2021}
{Nakariakov}, V.~M., {Anfinogentov}, S.~A., {Antolin}, P., {et~al.} 2021, \ssr,
  217, 73

\bibitem[{{Ni} {et~al.}(2022){Ni}, {Guo}, {Zhang}, {Chen}, {Fang}, \&
  {Chen}}]{ni2022}
{Ni}, Y.~W., {Guo}, J.~H., {Zhang}, Q.~M., {et~al.} 2022, \aap, 663, A31

\bibitem[{{Okamoto} {et~al.}(2004){Okamoto}, {Nakai}, {Keiyama}, {Narukage},
  {UeNo}, {Kitai}, {Kurokawa}, \& {Shibata}}]{okam2004}
{Okamoto}, T.~J., {Nakai}, H., {Keiyama}, A., {et~al.} 2004, \apj, 608, 1124

\bibitem[{{Parker}(1988)}]{park1988}
{Parker}, E.~N. 1988, \apj, 330, 474

\bibitem[{{Pelouze} {et~al.}(2022){Pelouze}, {Auch{\`e}re}, {Bocchialini},
  {Froment}, {Miki{\'c}}, {Soubri{\'e}}, \& {Voyeux}}]{pelo2022}
{Pelouze}, G., {Auch{\`e}re}, F., {Bocchialini}, K., {et~al.} 2022, \aap, 658,
  A71

\bibitem[{{Pontin} \& {Hornig}(2020)}]{pont2020}
{Pontin}, D.~I. \& {Hornig}, G. 2020, Living Reviews in Solar Physics, 17, 5

\bibitem[{{Ramsey} \& {Smith}(1966)}]{rams1966}
{Ramsey}, H.~E. \& {Smith}, S.~F. 1966, \aj, 71, 197

\bibitem[{{Roberts} {et~al.}(1984){Roberts}, {Edwin}, \& {Benz}}]{robe1984}
{Roberts}, B., {Edwin}, P.~M., \& {Benz}, A.~O. 1984, \apj, 279, 857

\bibitem[{{Sharma} {et~al.}(2022){Sharma}, {Oberoi}, {Battaglia}, \&
  {Krucker}}]{shar2022}
{Sharma}, R., {Oberoi}, D., {Battaglia}, M., \& {Krucker}, S. 2022, \apj, 937,
  99

\bibitem[{{Shen} {et~al.}(2014){Shen}, {Ichimoto}, {Ishii}, {Tian}, {Zhao}, \&
  {Shibata}}]{shen2014}
{Shen}, Y., {Ichimoto}, K., {Ishii}, T.~T., {et~al.} 2014, \apj, 786, 151

\bibitem[{{Su} {et~al.}(2012){Su}, {Wang}, {Veronig}, {Temmer}, \&
  {Gan}}]{su2012}
{Su}, Y., {Wang}, T., {Veronig}, A., {Temmer}, M., \& {Gan}, W. 2012, \apjl,
  756, L41

\bibitem[{{Tandberg-Hanssen}(1995)}]{tand1995}
{Tandberg-Hanssen}, E. 1995, {The nature of solar prominences}, Vol. 199

\bibitem[{{Tripathi} {et~al.}(2009){Tripathi}, {Isobe}, \& {Jain}}]{trip2009}
{Tripathi}, D., {Isobe}, H., \& {Jain}, R. 2009, \ssr, 149, 283

\bibitem[{{van Ballegooijen} {et~al.}(2014){van Ballegooijen}, {Asgari-Targhi},
  \& {Berger}}]{vanb2014}
{van Ballegooijen}, A.~A., {Asgari-Targhi}, M., \& {Berger}, M.~A. 2014, \apj,
  787, 87

\bibitem[{{Van Doorsselaere} {et~al.}(2020){Van Doorsselaere}, {Srivastava},
  {Antolin}, {Magyar}, {Vasheghani Farahani}, {Tian}, {Kolotkov}, {Ofman},
  {Guo}, {Arregui}, {De Moortel}, \& {Pascoe}}]{vand2020}
{Van Doorsselaere}, T., {Srivastava}, A.~K., {Antolin}, P., {et~al.} 2020,
  \ssr, 216, 140

\bibitem[{{Vernazza} {et~al.}(1981){Vernazza}, {Avrett}, \&
  {Loeser}}]{vern1981}
{Vernazza}, J.~E., {Avrett}, E.~H., \& {Loeser}, R. 1981, \apjs, 45, 635

\bibitem[{{Vial} \& {Engvold}(2015)}]{vial2015}
{Vial}, J.-C. \& {Engvold}, O. 2015, Astrophysics and Space Science Library,
  Vol. 415, {Solar Prominences}

\bibitem[{{Wedemeyer-B{\"o}hm} {et~al.}(2012){Wedemeyer-B{\"o}hm}, {Scullion},
  {Steiner}, {Rouppe van der Voort}, {de La Cruz Rodriguez}, {Fedun}, \&
  {Erd{\'e}lyi}}]{wede2012}
{Wedemeyer-B{\"o}hm}, S., {Scullion}, E., {Steiner}, O., {et~al.} 2012, \nat,
  486, 505

\bibitem[{{Withbroe} \& {Noyes}(1977)}]{with1977}
{Withbroe}, G.~L. \& {Noyes}, R.~W. 1977, \araa, 15, 363

\bibitem[{{Xia} {et~al.}(2012){Xia}, {Chen}, \& {Keppens}}]{xia2012}
{Xia}, C., {Chen}, P.~F., \& {Keppens}, R. 2012, \apjl, 748, L26

\bibitem[{{Xia} {et~al.}(2011){Xia}, {Chen}, {Keppens}, \& {van
  Marle}}]{xia2011}
{Xia}, C., {Chen}, P.~F., {Keppens}, R., \& {van Marle}, A.~J. 2011, \apj, 737,
  27

\bibitem[{{Xia} \& {Keppens}(2016)}]{xia2016}
{Xia}, C. \& {Keppens}, R. 2016, \apj, 823, 22

\bibitem[{{Xia} {et~al.}(2018){Xia}, {Teunissen}, {El Mellah}, {Chan{\'e}}, \&
  {Keppens}}]{xia2018}
{Xia}, C., {Teunissen}, J., {El Mellah}, I., {Chan{\'e}}, E., \& {Keppens}, R.
  2018, \apjs, 234, 30

\bibitem[{{Zhou} {et~al.}(2020){Zhou}, {Chen}, {Hong}, \& {Fang}}]{zhou2020}
{Zhou}, Y.~H., {Chen}, P.~F., {Hong}, J., \& {Fang}, C. 2020, Nature Astronomy,
  4, 994

\bibitem[{{Zhou} {et~al.}(2021){Zhou}, {Ruan}, {Xia}, \& {Keppens}}]{zhou2021}
{Zhou}, Y.-H., {Ruan}, W.-Z., {Xia}, C., \& {Keppens}, R. 2021, \aap, 648, A29

\bibitem[{{Zhou} {et~al.}(2018){Zhou}, {Xia}, {Keppens}, {Fang}, \&
  {Chen}}]{zhou2018}
{Zhou}, Y.-H., {Xia}, C., {Keppens}, R., {Fang}, C., \& {Chen}, P.~F. 2018,
  \apj, 856, 179

\bibitem[{{Zou} {et~al.}(2016){Zou}, {Fang}, {Chen}, {Yang}, {Hao}, \&
  {Cao}}]{zou2016}
{Zou}, P., {Fang}, C., {Chen}, P.~F., {et~al.} 2016, \apj, 831, 123

\end{thebibliography}

\clearpage

\begin{appendix}
\section{Ionization Degree in MPI-AMRVAC code}
\label{app1}
In the original MPI-AMRVAC 2.2 code \citep{xia2018}, the atmosphere is assumed to be fully ionized. To treat the H$\alpha$ emission in a more self-consistent way, we include ionization degree (of {hydrogen}) in our simulation. Since the method is not described either in our previous work \citep{zhou2020} or the MPI-AMRVAC paper \citep{xia2018}, we give the details in this Appendix.

In the MPI-AMRVAC code, conservative variables, for instance, density $\rho$, momentum $\rho v$, energy (volume) density $e$ and magnetic induction $B$, are solved. {The quantity $e$ is expressed as:} $e=e_{\mathrm{int}}+\rho v^2/2+B^2/2\mu_0$. {It} is composed of internal energy $e_{\mathrm{int}}$, kinetic energy $\rho v^2/2,$ and magnetic energy $B^2/2$.

When{} ionization is considered, we have the ionization fraction of $i=n_\mathrm{e}/n_\mathrm{H}$, where $n_\mathrm{H}=\rho/m_H$ is the total number density of {hydrogen}, $m_H$ is the mass of a {hydrogen} atom, and $n_\mathrm{e}$ is the number density of electron or ionized {hydrogen}. {The ionization fraction} $i$ is a function of vertical altitude $y$, thermal pressure $p$ and temperature $T$; for instance, we have $i=i \left(y,p,T\right)$, as tabulated in \citet{hein2015}.

Then, the internal energy $e_{\mathrm{int}}$ and gas pressure $p$ could be written as
  \begin{eqnarray}
    e_{\mathrm{int}} &=& \frac{p}{\gamma-1}+in_\mathrm{H}\chi,
    \label{eqa1}
    \\
    p &=& \left(n_\mathrm{e}+n_\mathrm{H}\right)k_BT=\left(1+i\right)n_\mathrm{H}k_\mathrm{B}T,
    \label{eqa2}
  \end{eqnarray}
respectively. $\gamma=5/3$ is the adiabatic index, $k_\mathrm{B}$ is the Boltzmann constant and $\chi=13.6$~eV is the ionization energy of {hydrogen}.

In the future 3.1 version of MPI-AMRVAC, the ionization degree, $i,$ is will be function of temperature only. Such $i(T)$ function has been tabulated, for instance, in \citet{carl2012} and \citet{hong2022}. Then, we can rewrite Eq.~\ref{eqa1} and Eq.~\ref{eqa2} as:
  \begin{equation}
    \frac{e_{\mathrm{int}}}{n_\mathrm{H}}=\frac{(1+i)k_\mathrm{B}T}{\gamma-1}+i\chi.
  \label{eqa3}
  \end{equation}
Considering that $i\left(T\right)$ is monotonically increasing with $T$, the right-hand side of Eq.~\ref{eqa3} is monotonic. Therefore, $T$ as well as $i$ could be determined from $e_{\mathrm{int}}/n_\mathrm{H}$, according to the dependence of $i\left(T\right)$. Therefore, after the update of $e$ and $\rho$ in the code, the variables $T$ and $i$ are then determined from $e_{\mathrm{int}}/n_\mathrm{H}$.

However, in the current work, $i$ is not a single function of $T$, but instead it is dependent on $y$, $p,$ and $T$.
We cannot use the same simple way to calculate $i$ and $T$ from $e_{\mathrm{int}}$ and $n_\mathrm{H}$.
For convenience, approximately, we can drop the ionization energy term $n_\mathrm{e}\chi$ on the right hand side of Eq.~\ref{eqa1}.
This approximation will overestimate the temperature a little bit especially when temperature is low, but still acceptable.
Then, at a certain grid point, with known $p$ and $y$, the ionization fraction, $i,$ is again a single function of temperature, $T$.
In this way, again, $i$ and $T$ could be obtained from $n_\mathrm{H}$ and $e_{\mathrm{int}}$ simultaneously.

Since the table given in \citet{hein2015} is sparse, with only 3-6 points on $y$, $p,$ and $T$, we used linear interpolation in between in our code.
For values beyond the table, we simply use the nearest value.
For example, the altitude $y$ is from 10~Mm to 30~Mm in the table.
Thus, for the region above 30~Mm, we used the value $i(y=30$~Mm$, p, T)$ directly, instead of extrapolating on the basis of the table.
As a result, the ionization fraction $i$ has a range from 0.17 to 0.94 in our simulation, instead of 0 to 1.
\end{appendix}

\end{document}